\documentclass[twocolumn,showpacs,preprintnumbers,amsmath,amssymb,aps,prb,longbibliography]{revtex4-2}
\usepackage{graphicx}
\usepackage{dcolumn}
\usepackage{bm}
\usepackage{verbatim} 
\usepackage[usenames, dvipsnames]{color} 
\usepackage{ulem}
\usepackage{color}
\usepackage[colorlinks=false,pdfencoding=auto,psdextra]{hyperref}
\usepackage{multirow}
\usepackage{tabularx}
\usepackage{dsfont}

\definecolor{orange}{rgb}{1.0, 0.5, 0.0}

\begin{document}
\title{ Weak dispersion of exciton Land{\'e} factor with band gap energy in lead halide perovskites: Approximate compensation of the electron and hole dependences} 


\author{N.~E. Kopteva$^{1}$, D.~R. Yakovlev$^{1,2}$, E. Kirstein$^{1}$, E.~A. Zhukov$^{1,2}$, D. Kudlacik$^{1}$, I.~V. Kalitukha$^{1,2}$, V.~F. Sapega$^{2}$, D.~N. Dirin$^{3}$,  M.~V. Kovalenko$^{3,4}$, A. Baumann$^{5}$, J. H\"ocker$^{5}$, V. Dyakonov$^{5}$, S.~A. Crooker$^{6}$, and M. Bayer$^{1}$}

\affiliation{$^{1}$Experimentelle Physik 2, Technische Universit\"at Dortmund, 44227 Dortmund, Germany}
\affiliation{$^{2}$Ioffe Institute, Russian Academy of Sciences, 194021 St.~Petersburg, Russia}
\affiliation{$^{3}$Department of Chemistry and Applied Biosciences, Laboratory of Inorganic Chemistry, ETH Zürich, 8093 Zürich, Switzerland}
\affiliation{$^{4}$Department of Advanced Materials and Surfaces, Laboratory for Thin Films and Photovoltaics, Empa—Swiss Federal Laboratories for Materials Science and Technology, 8600 Dübendorf, Switzerland}
\affiliation{$^{5}$Experimental Physics VI, Julius-Maximilian University of Würzburg, 97074 Würzburg, Germany}
\affiliation{$^{6}$National High Magnetic Field Laboratory, Los Alamos National Laboratory, Los Alamos, New Mexico 87545, USA}

\date{\today}

\begin{abstract}
The photovoltaic and optoelectronic properties of lead halide perovskite semiconductors are controlled by excitons, so that investigation of their fundamental properties is of critical importance. The exciton Land\'e or $g$-factor $g_\text{X}$ is the key parameter, determining the exciton Zeeman spin splitting  in magnetic fields. The exciton,  electron and hole carrier $g$-factors provide information on the band structure, including its anisotropy, and the parameters contributing to the electron and hole effective masses. We measure $g_\text{X}$ by reflectivity in magnetic fields up to 60~T for lead halide perovskite crystals. The materials  band gap energies at a liquid helium temperature vary widely across the visible spectral range from 1.520 up to 3.213~eV in hybrid organic-inorganic and fully inorganic perovskites with different cations and halogens: FA$_{0.9}$Cs$_{0.1}$PbI$_{2.8}$Br$_{0.2}$, MAPbI$_3$, FAPbBr$_3$, CsPbBr$_3$, and MAPb(Br$_{0.05}$Cl$_{0.95}$)$_3$. We find the exciton $g$-factors to be nearly constant, ranging from $+2.3$ to $+2.7$. Thus, the strong dependences of the electron and hole $g$-factors on the band gap roughly compensate each other when combining to the exciton $g$-factor. The same is true for the anisotropies of the carrier $g$-factors, resulting in a nearly isotropic exciton $g$-factor. The experimental data are compared favorably with model calculation results. 
\end{abstract}

\maketitle


\section{Introduction}

Lead halide perovskite semiconductors attract currently great attention due to their exceptional electronic and optical properties, which make them highly promising for applications in photovoltaics, optoelectronics, radiation detectors, etc.~\cite{Vardeny2022_book,Vinattieri2021_book, nrel2021, jena2019, nazarenko2017} Their flexible chemical composition $A$Pb$X_3$, where the cation $A$ can be cesium (Cs$^+$), methylammonium (MA$^+$), formamidinium (FA$^+$) and the anion $X$ can be I$^-$, Br$^-$, Cl$^-$, offers huge tunability of the band gap from the infrared up to the ultraviolet spectral range. 

The optical properties of perovskite semiconductors in vicinity of the band gap are controlled by excitons, which are electron-hole pairs bound by the Coulomb interaction. The exciton binding energy ranges from 14 to 64~meV~\cite{Galkowski2016,Baranowski2020a}, making them stable at room temperature at least for the large binding energies. In-depth studies of exciton properties such as of their energy and spin level structure or their relaxation dynamics, disclosing unifying trends for the whole class of lead halide perovskites, are therefore of key importance for basic and applied research.

The  band structure of lead halide perovskites is inverted compared to conventional III-V and II-VI semiconductors. As a result, the strong spin-orbit interaction influences the conduction rather than the valence band. Spin physics provides high precision tools for addressing electronic states in vicinity of the band gap: the Land\'e $g$-factors of electrons and holes are inherently linked via their values and anisotropies to the band parameters, which in turn determine the charge carrier effective masses~\cite{ivchenko2005,Yu2016}. On the other hand, the $g$-factors are the key parameters for the coupling of carrier spins to a magnetic field and thus govern spin-related phenomena and spintronics applications, a largely uncharted area for perovskites. 

We recently showed experimentally and theoretically that a universal dependence of the electron and hole $g$-factors showing strong variations with the band gap energy can be established for the whole family of hybrid organic-inorganic and fully inorganic lead halide perovskite crystals~\cite{kirstein2021nc}. These measurements were performed by time-resolved Kerr rotation and spin-flip Raman scattering spectroscopy including an analysis of the $g$-factor anisotropy. As the exciton $g$-factor $g_\text{X}$ is contributed by the electron and the hole $g$-factor, one may expect a similarly universal dependence for $g_\text{X}$. As some $g$-factor renormalization due to the finite carrier $k$-vectors in the exciton is expected, a direct measurement of the free exciton Zeeman splitting, e.g., by magneto-reflectivity or magneto-absorption, provides valuable insight. The data published so far concern mostly polycrystalline materials with broad exciton lines, which diminish the accuracy of exciton $g$-factor evaluation~\cite{Hirasawa1994,Tanaka2003,Galkowski2016,Baranowski2020a}. Magneto-optical studies of high quality crystals are needed to that end, in combination with high magnetic fields providing large Zeeman splittings~\cite{Yang2017,belykh2019,Baranowski2019}.

In this manuscript, we measured the exciton $g$-factors in lead halide perovskite crystals with different band gap energies using magneto-reflectivity in strong magnetic fields up to 60~T. We compare the experimental data with the electron and hole $g$-factors measured by time-resolved Kerr ellipticity and spin-flip Raman scattering.  We find that the exciton $g$-factor is nearly independent of the band gap energy that varies from 1.5 to 3.2~eV through the choice of cations and/or halogens in the perovskite composition. This behavior is in good agreement with the results of model calculations. We also find that the anisotropies of the electron and hole $g$-factors compensate each other, such that the net exciton $g$-factor becomes isotropic.


\section{Results}

We studied five lead halide perovskite crystals with band gap energies ($E_\text{g}$) covering basically the whole visible spectral range from 1.5 up to 3.2~eV. Details of the synthesis of these high-quality crystals are given in the Supporting Information, S1. The hybrid organic-inorganic compounds FA$_{0.9}$Cs$_{0.1}$PbI$_{2.8}$Br$_{0.2}$ ($E_\text{g} = 1.520$\,eV at the cryogenic temperature of $T = 1.6$~K) and MAPbI$_{3}$ ($E_\text{g} = 1.652$~eV) have band gap energies close to the near-infrared. Replacing the iodine halogen with bromine and chlorine results in a blue-shift of the band gap for FAPbBr$_{3}$ (2.216~eV) and MAPb(Br$_{0.05}$Cl$_{0.95}$)$_3$ (3.213~eV). To develop a complete picture, we also study the fully inorganic perovskite CsPbBr$_{3}$ (2.359~eV). 

\subsection{Optical properties}

\begin{figure*}[hbt]
\begin{center}
\includegraphics[width = 18cm]{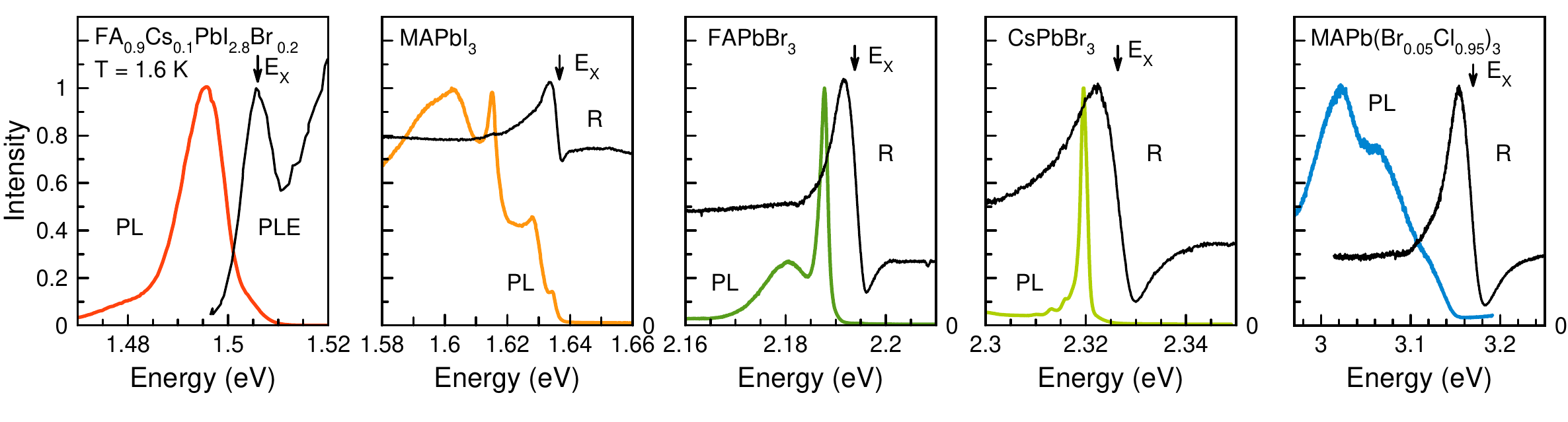}
\caption{\label{fig1} Optical properties of lead halide perovskite crystals with various band gaps at a temperature of 1.6~K. The colored lines show the photoluminescence spectra for continuous wave excitation at $E_\text{exc} = 3.06$~eV, using the laser power density of $5$~mW/cm$^2$.  For MAPb(Br$_{0.05}$Cl$_{0.95}$)$_3$ the excitation energy is $E_\text{exc} = 3.23$~eV. The black lines show the PLE spectrum for FA$_{0.9}$Cs$_{0.1}$PbI$_{2.8}$Br$_{0.2}$ and the reflectivity spectra for the other samples.}
\end{center}
\end{figure*}

An overview of the optical properties of the studied crystals at the temperature of $T = 1.6$~K is given in Figure~\ref{fig1}. We are interested in the properties of free excitons, which exhibit pronounced resonances in the reflectivity spectra of four studied crystals, but not for FA$_{0.9}$Cs$_{0.1}$PbI$_{2.8}$Br$_{0.2}$. For the latter, we measured the photoluminescence excitation (PLE) spectrum, where the pronounced peak at 1.506~eV corresponds to the exciton resonance energy, $E_\text{X}$. The exciton resonances are marked by the arrows in all panels and their energies are summarized in Table~\ref{tab:t1}. 

Photoluminescence (PL) spectra are also shown in Figure~\ref{fig1}. At low temperature, the excitons are typically localized or bound to impurities, and the emission from free excitons is weak. It can be seen as weak PL shoulder for FA$_{0.9}$Cs$_{0.1}$PbI$_{2.8}$Br$_{0.2}$ and MAPbI$_{3}$ only. The PL spectra are Stokes-shifted from the free exciton resonances and are composed of several lines, the origins of which are discussed in literature, but are not yet fully clarified~\cite{Chirvony2018,deQuilettes2019,Bercegol2018,kahmann2021}. The recombination dynamics, examples of which we show in the Supporting Information, Figure S1, are complex and show different characteristic time scales ranging from hundreds of picoseconds, as typical for free exciton lifetimes, up to tens of microseconds. The longer times are associated with resident electrons and holes, which are spatially separated due to localization at different sites. The dispersion in distances between an electron and a hole, resulting in different overlaps between their wave functions and trapping-detrapping processes, provide a strong variation of decay times. The coexistence of long-lived localized electrons and holes, which we refer to as resident carriers, is typical for lead halide perovskites, as confirmed by spin-dependent experimental techniques~\cite{belykh2019,kirstein2021nc,kirstein2021,kirstein2022mapi}.

\subsection{Measurement of exciton, electron and hole $g$-factors}

The band gap in lead halide perovskite semiconductors is located at the $R$-point of the Brillouin zone for the cubic crystal lattice and at the $\Gamma$-point for the tetragonal or orthorhombic lattices~\cite{Even2015}. In all these cases the states at the bottom of the conduction band and the top of the valence band have spin $1/2$. In an external magnetic field $B$, the electron (hole) Zeeman splittings, $E_\text{Z,e(h)} = \mu_\text{B} g_\text{e(h)} B$, are determined by the $g$-factors, $g_\text{e}$ and $g_\text{h}$. Here $\mu_\text{B}$ is the Bohr magneton. 

The bright (optically-active) exciton, composed of an electron and a hole, has the angular momentum $L =\pm 1$. Its spin sublevels are split by the Zeeman energy $E_\text{Z} = \mu_\text{B} g_\text{X} B$ with the exciton $g$-factor $g_\text{X}$. In the Faraday geometry with the magnetic field parallel to the light wave vector ($\textbf{B}_{\rm F} \parallel \textbf{k}$), the exciton states with opposite spin orientation can be distinguished by the circular polarization of the reflected light.

\begin{figure*}[hbt]
\begin{center}
\includegraphics[width = 12cm]{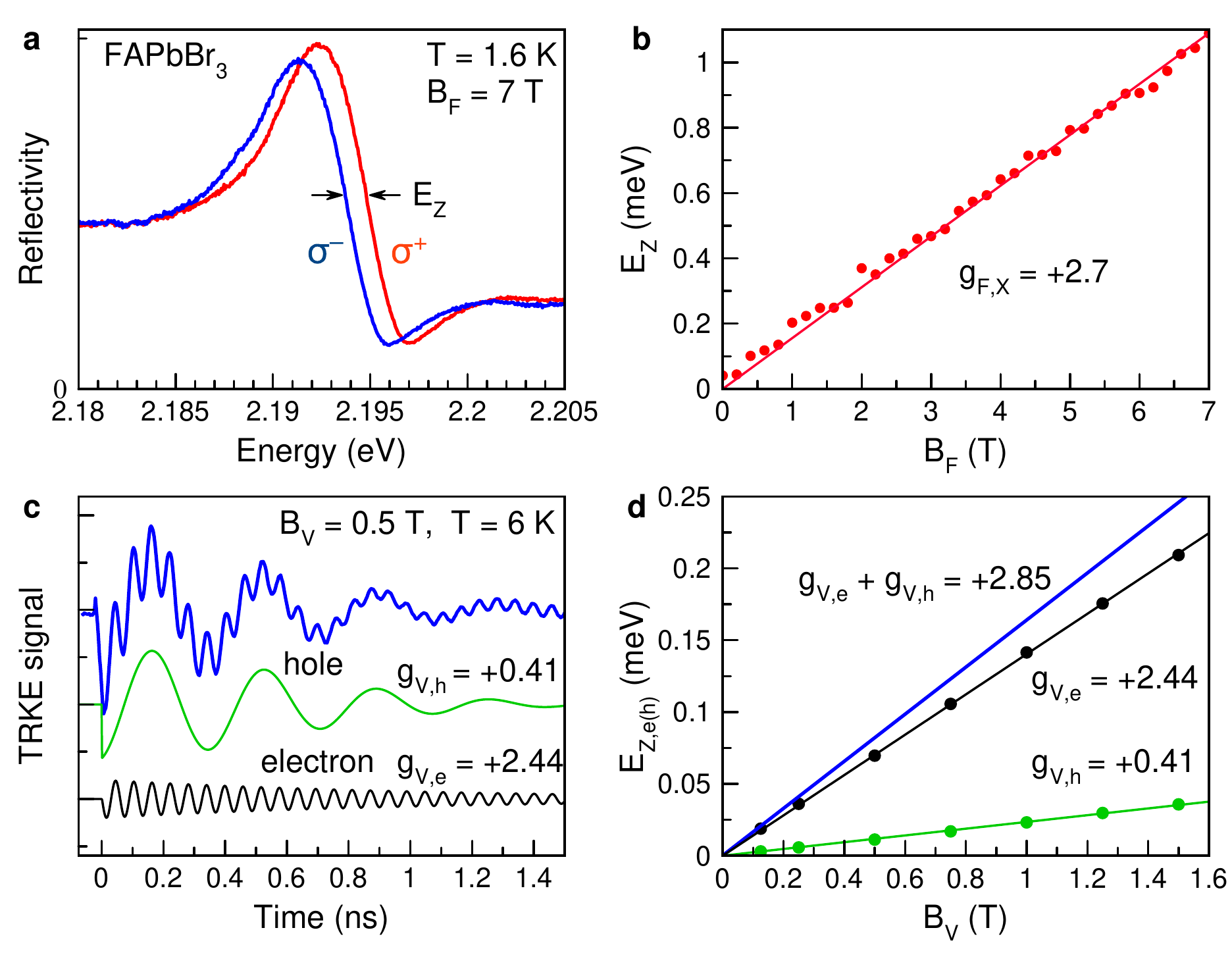}
\caption{\label{fig2} Magneto-optical properties of excitons as well as resident electrons and holes in a FAPbBr$_3$ crystal. (a) Reflectivity spectra measured in $\sigma^+$ (red line) and $\sigma^-$ (blue line) polarization in the longitudinal magnetic field $B_\text{F} = 7$\,T at $T = 1.6$\,K.  (b) Exciton Zeeman splitting as function of $B_\text{F}$ measured in magneto-reflectivity. Slope of the linear fit gives $g_\text{F,X} = +2.7$. (c) Time-resolved Kerr ellipticity signal (blue) measured at $B_\text{V} = 0.5$\,T, $T = 6$\,K, using the laser photon energy of $2.188$\,eV. The electron (black) and hole (green) components are obtained from decomposing the signal. (d) Dependence of the Zeeman splitting of the hole (green), the electron (black), and their sum (blue) on $B_\text{V}$.  
}
\end{center}
\end{figure*}

An example of such measurements for the FAPbBr$_3$ crystal is given in Figure~\ref{fig2}a, where the reflectivity spectra detected in $\sigma^+$ or $\sigma^-$ polarization at $B_\text{F}=7$~T are shown. One clearly sees the Zeeman splitting of the exciton resonance with $E_\text{Z}=1.09$~meV. The dependence of $E_\text{Z}$ on $B_\text{F}$ is a linear one, see Figure~\ref{fig2}b. From the slope we evaluate the exciton $g$-factor $g_\text{F,X} = +2.7$. Note that in this experiment the $g$-factor sign can be determined: positive values correspond to a high (low) energy shift of the $\sigma^+$ ($\sigma^-$) polarized resonance. We performed similar magneto-reflectivity experiments for MAPbI$_3$, CsPbBr$_3$, and MAPb(Br$_{0.05}$Cl$_{0.95}$)$_3$ and the measured exciton $g$-factors ($g_\text{F,X}$) are given in Table~\ref{tab:t1}.

The Zeeman splitting of the bright exciton is given by the sum of the electron and hole Zeeman splittings, so that 
\begin{equation}
\label{eq:gX}
g_\text{X}=g_\text{e}+g_\text{h}. 
\end{equation}
To what extent this equation is exactly fulfilled, needs to be checked specifically for each material, as some renormalization of the exciton $g$-factor may take place when the carriers are bound to form an exciton, in which they both are in motion. The renormalization could be caused by band mixing at finite wave vectors. 
We show below that Eq.~\eqref{eq:gX} is reasonably well fulfilled for the studied materials.   

Time-resolved Kerr rotation and Kerr ellipticity are powerful techniques for measuring directly the carrier $g$-factors by analyzing the Larmor precession frequencies, $\omega_{\rm L}$, of their spins~\cite{Yakovlev_Ch6}. In lead halide perovskites, resident electrons and holes coexist at low temperatures, so that their $g$-factors can be measured in one sample in a single experiment~\cite{belykh2019,kirstein2021nc,kirstein2021,kirstein2022mapi,Odenthal2017,chamarro2021}. We perform measurements of the time-resolved Kerr ellipticity (TRKE) on the FAPbBr$_3$ crystal, with the pump and probe laser energy tuned to the exciton resonance. The TRKE dynamics measured in the Voigt geometry with the magnetic field perpendicular to the light wave vector ($\textbf{B}_{\rm V} \perp \textbf{k}$, ${B}_{\rm V}=0.5$~T) are shown in Figure~\ref{fig2}c by the blue line. It contains two Larmor precession frequencies, which we decompose by the fit function given in the Experimental Section. We plot the electron and hole contributions separately in the same figure. Based on the established universal dependence of the carrier $g$-factors on the band gap energy in lead halide perovskites~\cite{kirstein2021nc}, we assign the higher Larmor frequency to the electrons and the smaller frequency to the holes. Their $g$-factors of $g_\text{V,e} = +2.44$ and $g_\text{V,h} = +0.41$ are evaluated with high accuracy from the magnetic field dependence of the Zeeman splitting $E_\text{Z,e(h)} = \hbar \omega_\text{L,e(h)}$ (Figure~\ref{fig2}d), where $\hbar$ is the reduced Planck constant, using $|g_{\rm e(h)}|=  \hbar \omega_{\rm L, e(h)}/ (\mu_{\rm B} B)$. 

The sign of the $g$-factor cannot be determined directly from the TRKE dynamics in the Voigt geometry. However, dynamic nuclear polarization in tilted magnetic field~\cite{kirstein2021} and theoretical calculation of the carrier $g$-factors' universal dependence~\cite{kirstein2021nc} allow us to identify it as positive for both electrons and holes in FAPbBr$_3$. 

We measure the anisotropy of the electron and hole $g$-factors in FAPbBr$_3$ by rotating the magnetic field using a vector magnet, see the Supporting Information, Figure S2. The $g$-factor anisotropy is small, namely the $g$-factors in the Faraday geometry ($g_\text{F,e} = +2.32$ and $g_\text{F,h} = +0.36$) are close to the ones in the Voigt geometry ($g_\text{V,e} = +2.44$ and $g_\text{V,h} = +0.41$). This is typical for lead halide perovskites with an FA cation, as in these materials the structural tolerance factor is close to unity and structural modifications at low temperatures are weak. Also for FA$_{0.9}$Cs$_{0.1}$PbI$_{2.8}$Br$_{0.2}$, we earlier reported nearly isotropic electron and hole $g$-factors~\cite{kirstein2021}. 

The sum of the carrier $g$-factors $g_\text{F,e} + g_\text{F,h} = +2.68$ in FAPbBr$_3$ coincides closely with the exciton $g$-factor $g_\text{F,X} = +2.7$ obtained from magneto-reflectivity measurements. These values are also close for MAPbI$_3$ ($g_\text{F,e} + g_\text{F,h} = +2.23$ and $g_\text{F,X} = +2.3$) and furthermore are not differ much in CsPbBr$_3$ ($g_\text{F,e} + g_\text{F,h} = +2.71$ and $g_\text{F,X} = +2.35$), see Table~\ref{tab:t1}. Note that the experimental accuracy  of $g$-factor values received from magneto-reflectivity is $\pm 0.1$ and    from the  TRKE is $\pm 0.05$.

\begin{table*}[thb!]
\caption{Parameters of lead halide perovskite crystals at cryogenic temperatures of $1.6-10~$~K.  
$^*$MAPbI$_3$ shows a complex anisotropy of the electron and hole $g$-factors~\cite{kirstein2021nc}. Here the given values of $g_\text{V}$ correspond to $\theta = 60^\circ$ and $g_\text{F}$ to $\theta = 150^\circ$. $^\#$The value is obtained from a linear approximation of the dependence of the exciton binding energy $E_\text{B}$ on $E_\text{g}$. $m_\text{X}$ is the exciton effective mass.  $^\&$In the original papers, the sign of the $g$-factor was not given.}
\label{tab:t1}
\begin{center}
\begin{tabular*}{1\textwidth}{@{\extracolsep{\fill}} |>{\centering\arraybackslash} m{0.17\textwidth}|>{\centering\arraybackslash} m{0.07\textwidth}|>{\centering\arraybackslash} m{0.07\textwidth}|>{\centering\arraybackslash} m{0.05\textwidth}|>{\centering\arraybackslash} m{0.05\textwidth}|>{\centering\arraybackslash} m{0.05\textwidth}|>{\centering\arraybackslash} m{0.05\textwidth}|>{\centering\arraybackslash} m{0.08\textwidth}|>{\centering\arraybackslash} m{0.05\textwidth}|>{\centering\arraybackslash} m{0.05\textwidth}|>{\centering\arraybackslash} m{0.08\textwidth}|>{\centering\arraybackslash} m{0.09\textwidth}|}
\hline
Material & $E_\text{g}$ (eV) & $E_\text{X}$ (eV)& $E_\text{B}$ (meV)& $m_\text{X}$ ($m_0$)~\cite{Galkowski2016} & $g_\text{V,e}$~\cite{kirstein2021nc} & $g_\text{V,h}$~\cite{kirstein2021nc} & $g_\text{V,e} + g_\text{V,h}$ & $g_\text{F,e}$~\cite{kirstein2021nc} & $g_\text{F,h}$~\cite{kirstein2021nc}& $g_\text{F,e} + g_\text{F,h}$ & $g_\text{F,X}$\\
\hline
FA$_{0.9}$Cs$_{0.1}$PbI$_{2.8}$Br$_{0.2}$ & 1.520 & 1.506 & 14~\cite{Galkowski2016} & 0.09 & +3.48 & -1.15 & +2.33 & +3.72 & -1.29 & +2.43 & - \\
\hline
MAPbI$_3$ ($^*$) & 1.652 & 1.636 & 16~\cite{Galkowski2016} & 0.104 & +2.81 & -0.68 & +2.13 & +2.57 & -0.34 & +2.23 & +2.3\\
\hline
MAPbI$_3$~\cite{Yang2017} & 1.656 & 1.640 & 16~\cite{Galkowski2016} & 0.104 & - & - & - & - & - & - & $2.66 \pm 0.1^\&$ \\
\hline
FAPbBr$_3$ & 2.216 & 2.194 & 22~\cite{Galkowski2016} & 0.115 & +2.44  & +0.41 & +2.85 & +2.32 & +0.36 & +2.68 & +2.7\\
\hline
MAPbBr$_3$~\cite{Baranowski2019} & 2.275 & 2.250 & 25~\cite{Baranowski2020a} & 0.117 & -  & - & - & - & - & - & 2.6$^\&$\\
\hline
CsPbBr$_3$ & 2.359 & 2.326 & 33~\cite{Yang2017b} & 0.126 & +1.69  & +0.85 &  +2.54 & +2.06 & +0.65 &  +2.71 & +2.35\\
\hline
MAPb(Br$_{0.05}$Cl$_{0.95}$)$_3$ & 3.213 & 3.168 & 45$^\#$ & - & - & +1.33 & - & - & - & - & +2.5\\
\hline
\end{tabular*}
\end{center}
\end{table*}

\subsection{Anisotropy of exciton $g$-factor}

\begin{figure}[b!]
\begin{center}
\includegraphics[width = 8.6cm]{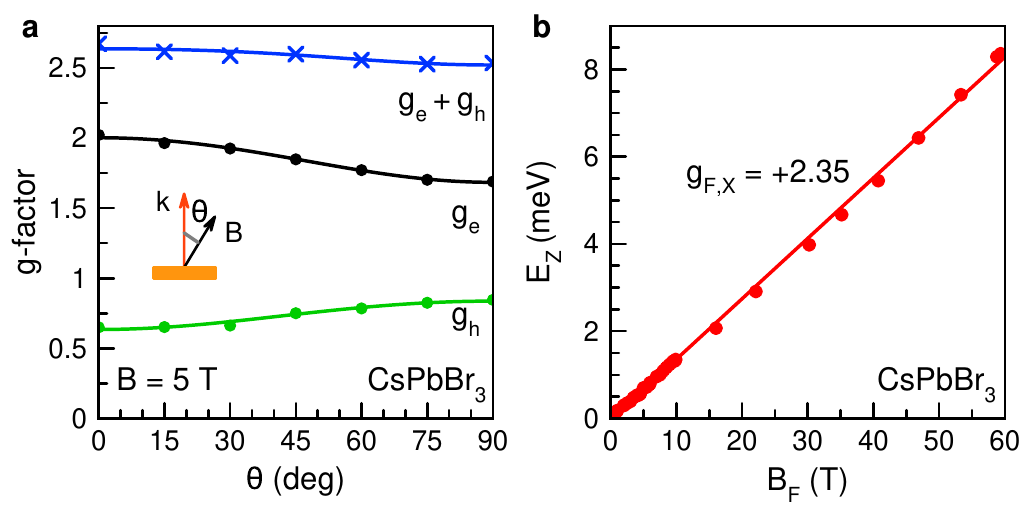}
\caption{\label{fig3}  
(a) Anisotropy of the electron (black circles) and hole (green circles) $g$-factors measured by SFRS for CsPbBr$_3$. $B = 5$\,T and $T = 1.6$\,K. The black and green lines are calculated with Eq.~\eqref{eq:angle}, using the parameters from Table~\ref{tab:t1}. The blue crosses are experimental data of $g_\text{e}+g_\text{h}$, the blue line is the sum of the fits shown by the black and green lines. (b) Magnetic field dependence of the exciton Zeeman splitting measured from magneto-reflectivity in the Faraday geometry for CsPbBr$_3$. $T = 1.6$~K. The symbols are experimental data and the line is a linear fit.}
\end{center}
\end{figure}

In CsPbBr$_3$ crystals, the anisotropy of the electron and hole $g$-factors is pronounced. The results measured by spin-flip Raman scattering (SFRS), for details see the Supporting Information, Figure S3, are shown in Figure~\ref{fig3}a. Here $\theta$ is the angle of the magnetic field tilt from the Faraday ($\theta=0^\circ$) to the Voigt ($\theta=90^\circ$) geometry. In this experiment, the c-axis is oriented perpendicular to the light $k$-vector, so that $\textbf{B} \parallel c$ corresponds to the Voigt geometry.  Note, that contrary to TRKE, spin-flip lines are detectable also in the Faraday geometry and, therefore, $g_\text{F,e(h)}$ are measured directly. The $g_\text{e}$ and $g_\text{h}$ dependences on $\theta$ are described by 
\begin{equation}
\label{eq:angle}
g_\text{e(h)}(\theta) = \sqrt{g_\text{F,e(h)}^2\cos^2\theta + g_\text{V,e(h)}^2\sin^2\theta}. 
\end{equation}
The electron and hole $g$-factors are both positive in CsPbBr$_3$, but their anisotropies are orthogonal to each other. The electron $g$-factor is largest in the Faraday geometry ($g_\text{F,e}=+2.06$) and decreases in the Voigt geometry to $g_\text{V,e}=+1.69$, while the hole $g$-factor increases from the Faraday towards the Voigt geometry from $g_\text{F,h}=+0.65$ up to $g_\text{V,h}=+0.85$. Interestingly, the sum of the carrier $g$-factors changes only weakly from $+2.71$ to $+2.54$, see the blue crosses in Figure~\ref{fig3}a, demonstrating that the exciton $g$-factor anisotropy is weak. This is a common feature for the lead halide perovskites, as confirmed by the data for MAPbI$_3$, where the strong anisotropies of the electron and hole $g$-factors nearly compensate each other in their sum, see Table~\ref{tab:t1}.  

\subsection{Exciton Zeeman splitting in strong pulsed magnetic fields}

We examined the exciton Zeeman splitting in the CsPbBr$_3$ crystal in very strong magnetic fields up to 60~T, using a pulsed magnet at the National High Magnetic Field Laboratory in Los Alamos. The experiments were motivated by the possibility to reach large Zeeman splittings, thus improving the accuracy of the exciton $g$-factor evaluation, and also by searching for a possible nonlinearity of the Zeeman splitting by field-induced band mixing. Magneto-reflectivity was measured at $T=1.6$~K and the Zeeman splitting of the oppositely circularly polarized exciton resonances (similar as in Figure~\ref{fig2}a) was assessed. The results are shown in Figure~\ref{fig3}b. The exciton Zeeman splitting increases linearly with magnetic field over the whole range up to 60~T. The evaluated $g_\text{F,X}=+2.35$. The high linearity indicates that band mixing is negligibly small in lead halide perovskites even in very strong magnetic fields. This is explained by the simple spin structure (spin 1/2) of the electronic states in the vicinity of the band gap, which contribute to the exciton wave function. The shift of the higher-lying electron states with momentum 3/2 due to the spin-orbit splitting in the conduction band is about 1.5~eV, which is large enough to exclude a significant admixture of these states to the ground exciton state by the magnetic field. Note that in this respect the lead halide perovskites principally differ from conventional III-V and II-VI semiconductors, for which strongly nonlinear Zeeman splittings were reported in GaAs- and CdTe-based quantum wells~\cite{Traynor1997,Bartsch2011}.   

\subsection{Band gap dependence of exciton $g$-factor}

\begin{figure}[hbt]
\begin{center}
\includegraphics[width = 8.5cm]{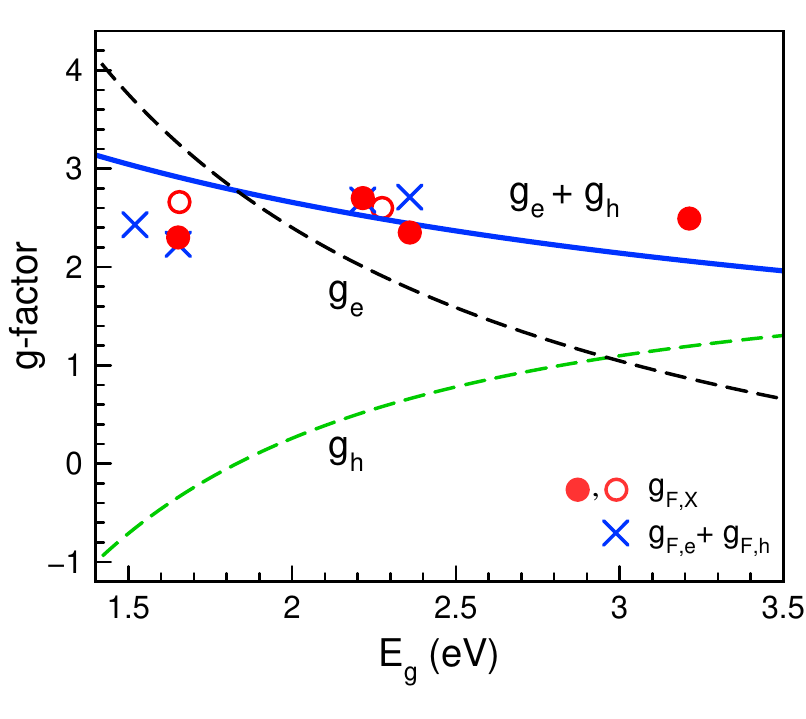}
\caption{\label{fig4} Dependence of the exciton $g$-factor measured by magneto-reflectivity (closed red circles are our data and open red circles are from Refs.~\cite{Yang2017,Baranowski2019}) and of the sum of carrier $g$-factors evaluated from TRKE and SFRS (crosses) on the band gap energy. The data are taken at cryogenic temperatures of $1.6-10$~K. The dashed lines are model calculations~\cite{kirstein2021nc} with Eqs.~\eqref{eq:gh} and \eqref{eq:ge} for the hole (green) and the electron (black) $g$-factors, which closely match the experimental data given in Table~\ref{tab:t1}. The blue line is the sum of their contributions calculated with Eq.~\eqref{eq:gx1}.
}
\end{center}
\end{figure}

To summarize the information on the exciton $g$-factors for the whole class of lead halide perovskites and highlight general trends, we show in Figure~\ref{fig4} the experimental data collected in Table~\ref{tab:t1} as function of the band gap energy. Our data for $g_\text{F,X}$ measured by magneto-reflectivity are shown by the closed red circles. We combine them with literature data (open red circles), also measured by magneto-reflectivity on MAPbI$_3$ and MAPbBr$_3$ crystals~\cite{Yang2017,Baranowski2019}. One can see that the exciton $g$-factor is nearly independent of the band gap energy varying from 1.5 to 3.2~eV. The exciton $g$-factors change only in a small range from $+2.3$ to $+2.7$. Compared to the average, this corresponds to a variation well below 10\%, while the electron $g$-factor varies by significantly more than 50\%.

We show in Figure~\ref{fig4} also the sum values of $g_\text{F,e}+g_\text{F,h}$ by the crosses, which follow closely the $g_\text{F,X}$ values. Therefore, we conclude that the renormalization of the carrier $g$-factors in the excitons is small in the lead halide perovskites. 

Let us compare our experimental data with model predictions. We showed recently experimentally and theoretically that the electron and hole $g$-factors in the lead halide perovskites follow universal dependences on the band gap energy~\cite{kirstein2021nc}. According calculations are given in Figure~\ref{fig4} by the dashed lines. They account for the fact that in the vicinity of the band gap the band structure of hybrid organic-inorganic and of fully inorganic lead halide perovskites is strongly contributed by Pb orbitals. Then, for the holes in the valence band, the main contribution to the $g$-factor is due to $\mathbf{k}\cdot\mathbf{p}$ mixing with the conduction band. For the cubic phase it is described by~\cite{kirstein2021nc,arellano2022}:
\begin{equation}
\label{eq:gh}
g_\text{h} = 2 - \frac{4}{3} \frac{p^2}{m_0}\left(\frac{1}{E_\text{g}} - \frac{1}{E_\text{g} + \Delta}\right).
\end{equation}
Here $p$ is the interband matrix element of the momentum operator, $\Delta = 1.5$\,eV is the spin-orbit splitting of the conduction band, $m_0$ is the electron effective mass, and $\hbar p/m_0 = 6.8$\,eV{\AA}. For electrons, the $\mathbf{k}\cdot\mathbf{p}$ mixing with the top valence band and the remote valence states are important: 
\begin{equation}
\label{eq:ge}
g_\text{e} = -\frac{2}{3} + \frac{4}{3} \frac{p^2}{m_0 E_\text{g}} + \Delta g_\text{e}.
\end{equation}
Here $\Delta g_\text{e} = -1$ is the remote band contribution evaluated as fitting parameter in Ref.~\cite{kirstein2021nc}. One clearly sees that both electron and hole $g$-factors change strongly with increasing $E_{\rm g}$. 

Using Eqs.~(\ref{eq:gh}) and (\ref{eq:ge}) we obtain for the sum of the electron and hole $g$-factors:
\begin{equation}
\label{eq:gx1}
g_\text{e} + g_\text{h} = \frac{4}{3} + \frac{4}{3}\frac{p^2}{m_0}\frac{1}{E_\text{g} + \Delta} + \Delta g_\text{e}.
\end{equation}
It is shown by the solid blue line in Figure~\ref{fig4}. One can see that it has much weaker dependence on the band gap compared to the $g$-factors of the individual carriers. The calculated dependence for $g_\text{e} + g_\text{h}$ is in good agreement with the experimental data on both $g_\text{F,X}$ and $g_\text{e} + g_\text{h}$ shown in Figure~\ref{fig4}.  Note that the similar cancellation is valid for the angle dependence of the exciton $g$-factor, shown exemplary for CsPbBr$_3$ in Figure~\ref{fig3}a.  The related formalism has been developed in Ref.~\cite{kirstein2021nc}, and we give its key equations in the Supporting Information, S5.

The observed, approximate independence of the exciton $g$ factor on the band gap $E_\text{g}$ is a consequence of the underlying simple band structure, determined by the lowest conduction band $c,\pm 1/2$ and the highest valence band $v, \pm 1/2$, showing both only a two-fold spin degeneracy. As shown above~\cite{kirstein2021nc}, the contributions to the individual $g$-factors from the $\mathbf{k}\cdot\mathbf{p}$ mixing of these bands proportional to $1/E_\text{g}$ cancel each other in the exciton $g$-factor. This is in striking contrast to the Zeeman splitting of the exciton sublevels in conventional III-V or II-VI bulk semiconductors of cubic symmetry, like GaAs or CdTe. First, in these well-studied semiconducting materials, the conduction band $\Gamma_6$ is simple, but the electron $g$ factor is strongly dependent not only on $E_\text{g}$, but also on the spin-orbit splitting $\Delta$ of the valence band $\Gamma_8$, the latter varying in a wide range~\cite{Roth1959}. Second and more importantly, the hole is characterized by the effective angular momentum $j=3/2$ so that it is four-fold degenerate. The valence band degeneracy leads to complicated exciton level splitting and structure of its wave function as result of the interplay of heavy hole-light hole mixing and electron-hole exchange interaction~\cite{Altarelli1973,Cho1975,Venghaus1979}. Even if the field-induced hole mixing and the exchange interaction in the exciton are disregarded, the effective $g$-factor of the bright exciton (A-exciton) contains $1/E_\text{g}$ contributions. 

While the mixing of the bands forming the gap cancels thus in $g_\text{X}$, the mixing of the valence band with the higher lying, remote conduction band still contributes, but here the variation with $E_\text{g}$ is damped by adding the spin-orbit splitting which is comparatively large in the perovskites. 
The situation  is similar to the one in transition metal dichalcogenide monolayers, where the bright exciton $g$-factor is as well largely determined by the remote bands, while the mutual contributions of the valence band to the electron $g$ factor and of the conduction band to the hole $g$-factor cancel out in the exciton $g$-factor~\cite{Wang2015,Wozniak2020,Deilmann2020}.

Up to now we discussed the band gap dependence of the bright exciton $g$-factor, $g_\text{X}  = g_\text{e} + g_\text{h}$. For the optically-forbidden dark exciton the $g$-factor value is given by the difference $g_\text{e} - g_\text{h}$. Due to the opposite variation of the electron and hole $g$-factors, the dependence on $E_\text{g}$ is expected to be even stronger for the dark exciton than for individual electron and hole, see the Supporting Information, Figure~S4. Also a stronger anisotropy for the dark exciton $g$-factor, compared to the bright exciton $g$-factor which is nearly isotropic, is expected. 

To summarize, the exciton $g$-factors in hybrid organic-inorganic and fully inorganic lead halide perovskites have positive signs and show relatively constant values over a large range of band gap energies. Further, they are nearly isotropic. It turns out that the strong band gap dependences and anisotropies of the individual electron and hole $g$-factors contributing to the exciton largely compensate each other. It would be important to extend these studies to perovskite semiconductors based on another metal ions, e.g., lead-free Sn-based materials. Also the role of quantum confinement for carriers and excitons in perovskite nanocrystals and two-dimensional materials would be interesting to examine.


\section{Experimental Section}

\textbf{Samples:} 
The class of lead halide perovskites possesses the composition $A$Pb$X_3$, where the $A$-cation is typically Cs$^+$, methylammonium (MA$^+$, CH$_3$NH$_3^+$), or formamidinium (FA$^+$, CH(NH$_2$)$_2^+$), and the $X$-anion is one of the halogens Cl$^-$, Br$^-$, or I$^-$, giving rise to a high flexibility. The flexibility is only limited by a favorable ratio of the anion to cation ion radii, named the Goldschmidt tolerance factor $t$, which should be close to unity~\cite{goldschmidt_gesetze_1926}. By varying the composition, the band gap of these perovskite materials can be tuned from the infrared up to the ultraviolet spectral range. All studied samples are lead halide perovskite single crystals grown out of solution with the inverse temperature crystallization (ITC) technique~\cite{Dirin2016,nazarenko2017,hocker2021}. For specific crystals the ITC protocols were modified, and for the five crystals studied here (FA$_{0.9}$Cs$_{0.1}$PbI$_{2.8}$Br$_{0.2}$, MAPbI$_3$, FAPbBr$_3$, CsPbBr$_3$, and MAPb(Br$_{0.05}$Cl$_{0.95}$)$_3$) details of their synthesis are given in the Supporting Information (Section S1) and in Ref.~\cite{kirstein2021nc}.

\textbf{Magneto-reflectivity:}
The samples were placed in a cryostat and immersed in either superfluid liquid helium at $T=1.6$~K or in gas helium for $T=6-10$~K. For experiments in \textit{dc} magnetic fields, two cryostats equipped with a split-coil superconducting magnet were used. One with a single split coil can generate magnetic fields up to 10~T in a fixed direction. Another is a vector magnet with three orthogonally oriented split coils, allowing us to apply magnetic fields up to 3~T in any direction. A sketch of the experimental geometry is shown in Figure~\ref{fig3}a. 

A halogen lamp was used for measuring magneto-reflectivity. The light wave vector, $\mathbf{k}$, was perpendicular to the sample surface and the reflectivity was measured in backscattering geometry. The signal was analyzed for $\sigma^+$ and $\sigma^-$ circular polarization and recorded after dispersion with an 0.5-meter spectrometer with a Silicon charge-coupled-device camera. The external magnetic field was applied parallel to the light $k$-vector, $\mathbf{B}_\text{F} \parallel \mathbf{k}$ (Faraday geometry). In the TRKE and SFRS experiments, we also used the Voigt geometry ($\mathbf{B}_\text{V} \perp \mathbf{k}$), as well as tilted geometries between Faraday and Voigt. Here the tilt angle $\theta$ is defined as the angle between $\mathbf{B}_\text{F}$ and $\mathbf{B}_\text{V}$, where $\theta = 0^\circ$ corresponds to the Faraday geometry. Experiments in pulsed magnetic fields up to 60~T were performed in the Faraday geometry, following previously-described methods~\cite{Stier:2016}.

\textbf{Time-resolved Kerr ellipticity (TRKE):}
The coherent spin dynamics were measured by a pump-probe setup, where pump and probe had the same photon energy, emitted from a pulsed laser~\cite{Yakovlev_Ch6}. The used titanium-sapphire (Ti:Sa) laser emitted 1.5~ps pulses with a spectral width of about $1$~nm (1.5~meV) at a pulse repetition rate of 76~MHz (repetition period $T_\text{R}=13.2$~ns). An optical parametric oscillator (OPO) with internal frequency doubling was used to convert the photon energy of the Ti:Sa laser so that it can be resonantly tuned to the exciton resonance in FAPbBr$_3$. 

The laser beam was split into two beams, the pump and the probe. The probe pulses were delayed with respect to the pump pulses by a mechanical delay line.  Both pump and probe beams were modulated using photo-elastic modulators (PEMs). The linear polarization of the probe was fixed, but its amplitude was modulated at the frequency of 84~kHz. The pump beam helicity was modulated between $\sigma^+$ and $\sigma^-$ circular polarization at the frequency of 50~kHz.  The elliptical polarization of the reflected probe beam was analyzed via balanced photodiodes and a lock-in amplifier (Kerr ellipticity). In a transverse magnetic field, the Kerr ellipticity amplitude oscillates in time due to the Larmor spin precession of the carriers, decaying at longer time delays. When both electrons and holes contribute to the Kerr ellipticity signal, which is the case for the studied perovskite crystals, the signal can be described as a superposition of two decaying oscillatory functions: $A_{\rm KE} = S_{\rm e} \cos (\omega_{\rm L, e} t) \exp(-t/T^*_{\rm 2,e}) + S_{\rm h} \cos (\omega_{\rm L, h} t) \exp(-t/T^*_{\rm 2,h})$. Here $S_{\rm e(h)}$ are the signal amplitudes that are proportional to the spin polarization of electrons (holes), and $T^*_{\rm 2,e(h)}$ are the carrier spin dephasing times. The $g$-factors are evaluated from the Larmor precession frequencies $\omega_{\rm L, e(h)}$ via $|g_{\rm e(h)}|=  \hbar \omega_{\rm L, e(h)}/ (\mu_{\rm B} B)$. It is important to note that TRKE provides information on the $g$-factor magnitude, but not on its sign. The sign of the carrier $g$-factors can be identified in tilted magnetic field geometry through the dynamic nuclear polarization effect, for details see Refs.~\cite{kirstein2021,kirstein2022mapi}. 


\textbf{Spin-flip Raman scattering (SFRS):} This technique allows us to measure directly the Zeeman splitting of the electron and hole spin sublevels via the spectral shift of the scattered light from the laser photon energy~\cite{hafele_chapter_1991}. The energy shift is provided by a spin-flip of the carriers, with the required energy taken from or provided by phonons. The typical shifts do not exceed 1~meV at magnetic field of 10~T, which demands a high spectral resolution provided by high-end spectrometers with excellent suppression of scattered laser light. The experiments were performed for the CsPbBr$_3$ sample immersed in superfluid liquid helium ($T=1.6$~K). We used resonant excitation at 2.330~eV in the vicinity of the exciton resonance of CsPbBr$_3$ in order to enhance the SFRS signal. The resonant Raman spectra were measured in backscattering geometry using the laser power density of 1~Wcm$^{-2}$. The scattered light was analyzed by a Jobin-Yvon U1000 double monochromator (1 meter focal length) equipped with a cooled GaAs photomultiplier and conventional photon counting electronics. The spectral resolution of 0.2~cm$^{-1}$ (0.024~meV) allowed us to measure the SFRS signals in close vicinity of the laser line for spectral shifts ranging from 0.1 to 3~meV. The SFRS spectra were measured for cross-polarized ($\sigma^- / \sigma^+$) circular polarizations of excitation ($\sigma^-$) and detection ($\sigma^+$).




\subsection*{Supporting Information}
Supporting Information is available from the Wiley Online Library or from the authors.

\subsection*{Acknowledgements}
The authors are thankful to M.~M.~Glazov, M.~O.~Nestoklon and E. L. Ivchenko for fruitful discussions. We acknowledge financial support by the Deutsche Forschungsgemeinschaft via the SPP2196 Priority Program (Project YA 65/26-1) and the International Collaborative Research Centre TRR160 (Projects A1 and B2). I.V.K. acknowledges support of the Russian Foundation for Basic Research (Project No. 19-52-12064).  The work at ETH Z\"urich (D.N.D. and M.V.K.) was financially supported by the Swiss National Science Foundation (grant agreement 186406, funded in conjunction with SPP219 through DFG-SNSF bilateral program) and by ETH Z\"urich through ETH+ Project SynMatLab. J.H., A.B. and V.D. acknowledge financial support from the Deutsche Forschungsgemeinschaft  through the W\"urzburg-Dresden Cluster of Excellence on Complexity and Topology in Quantum Matter-ct.qmat (EXC 2147, project-id 39085490), DY18/19-1 and from the Bavarian State Ministry of Education and Culture, Science and Arts within the Collaborative Research Network "Solar Technologies go Hybrid". Experiments  at the National High Magnetic Field Laboratory were supported by the National Science Foundation DMR-1644779, the State of Florida, and the US Department of Energy. 

\subsection*{Conflict of interest}
Authors declare no conflict of interest.

\subsection*{Data Availability Statement} 
The data on which the plots in this paper are based and other findings of our study are available from the corresponding authors upon justified request. 

\subsection*{Keywords}
Lead halide perovskites, spintronics, exciton Land\'e factor, electron $g$-factor, hole $g$-factor,  magneto-reflectivity, pump-probe Kerr ellipticity, spin-flip Raman scattering

\subsection*{Author information}
Correspondence should be addressed to N.E.K. (email: natalia.kopteva@tu-dortmund.de) and D.R.Y. (email: dmitri.yakovlev@tu-dortmund.de) \\
\\

\textbf{ORCID} \\
Natalia E. Kopteva:   0000-0003-0865-0393 \\
Dmitri R. Yakovlev:   0000-0001-7349-2745 \\
Erik Kirstein:        0000-0002-2549-2115 \\
Evgeny A. Zhukov:     0000-0003-0695-0093 \\
Dennis Kudlacik:      0000-0001-5473-8383\\
Ina V. Kalitukha:     0000-0003-2153-6667\\
Victor F. Sapega:     0000-0003-3944-7443\\
Dmitry N. Dirin:      0000-0002-5187-4555\\
Maksym V. Kovalenko:  0000-0002-6396-8938 \\
Andreas Baumann:      0000-0002-9440-0456 \\
Julian H\"ocker:      0000-0002-8699-3431\\
Vladimir Dyakonov:    0000-0001-8725-9573\\
Scott A. Crooker:     0000-0001-7553-4718\\
Manfred Bayer:        0000-0002-0893-5949\\

\end{document}


\title{Supporting Information: \\  Weak dispersion of exciton Land{\'e} factor with band gap energy in lead halide perovskites: Approximate compensation of the electron and hole dependences}

\author{N. E. Kopteva$^{1}$, D.~R. Yakovlev$^{1,2}$, E. Kirstein$^{1}$, E.~A. Zhukov$^{1,2}$, D. Kudlacik$^{1}$, I. V. Kalitukha$^{1,2}$, V. F. Sapega$^{2}$, D. N. Dirin$^{3}$,  M. V. Kovalenko$^{3,4}$, A. Baumann$^{5}$, J. Höcker$^{5}$, V. Dyakonov$^{5}$, S.~A.~Crooker$^{6}$, and M. Bayer$^{1}$}

\affiliation{$^{1}$Experimentelle Physik 2, Technische Universit\"at Dortmund, 44227 Dortmund, Germany}
\affiliation{$^{2}$Ioffe Institute, Russian Academy of Sciences, 194021 St. Petersburg, Russia}
\affiliation{$^{3}$Department of Chemistry and Applied Biosciences, Laboratory of Inorganic Chemistry, ETH Z\"urich, 8093 Z\"urich, Switzerland}
\affiliation{$^{4}$Department of Advanced Materials and Surfaces, Laboratory for Thin Films and Photovoltaics, Empa—Swiss Federal Laboratories for Materials Science and Technology, 8600 D\"ubendorf, Switzerland}
\affiliation{$^{5}$Experimental Physics VI, Julius-Maximilian University of W\"urzburg, 97074 W\"urzburg, Germany}
\affiliation{$^{6}$National High Magnetic Field Laboratory, Los Alamos National Laboratory, Los Alamos, New Mexico 87545, USA}

\date{\today}

\maketitle

\subsection*{S1. Samples}
The class of lead halide perovskites possesses $A$Pb$X_3$ composition, where the $A$-cation is typically Cs, methylammonium (MA$^+$, CH$_3$NH$_3^+$) or formamidinium (FA$^+$, CH(NH$_2$)$_2^+$), and the $X$-anion is one of the halogen ion Cl$^-$, Br$^-$, or I$^-$, giving rise to a high flexibility. This flexibility is only limited by a favorable ratio of the anion to cation ion radii, named the Goldschmidt tolerance factor $t$, which should be close to unity~\cite{goldschmidt_gesetze_1926}. By varying the composition, the band gap of these perovskite materials can be tuned from the infrared up to the ultraviolet spectral range. All studied samples are lead halide perovskite single crystals grown out of solution with the inverse temperature crystallization (ITC) technique~\cite{Dirin2016,nazarenko2017,hocker2021}. For specific crystals the ITC protocols were modified.     
 
\textbf{FA$_{0.9}$Cs$_{0.1}$PbI$_{2.8}$Br$_{0.2}$ crystals.}
$\alpha$-phase FA$_{0.9}$Cs$_{0.1}$PbI$_{2.8}$Br$_{0.2}$ single crystals were grown by the method described in detail in Ref.~\onlinecite{nazarenko2017}. First, a solution of CsI, FAI, PbI$_2$, and PbBr$_2$, with $\gamma$-butyrolactone (GBL) as solvent is mixed. This solution is then filtered and slowly heated to 130$^\circ$C temperature, whereby single crystals are formed in the black phase of FA$_{0.9}$Cs$_{0.1}$PbI$_{2.8}$Br$_{0.2}$. Afterward the crystals are separated by filtering and drying. The $\alpha$-phase (black phase) exhibits a cubic crystal structure at room temperature~\cite{weller_cubic_2015}. In the experiment the crystal was oriented with $[001]$ pointing along the laser wave vector $\textbf{k}$. Note that the $g$-factor isotropy, the small shift of the PL line with temperature, and further analysis \cite{nazarenko2017,kirstein2021} suggest that the typical lead halide perovskite crystal distortion from cubic symmetry is small at cryogenic temperatures. The size of the studied crystal is about $2 \times 3\times 2$~mm$^3$. The crystal shape is non-cuboid, but the crystal structure exhibits aristotype cubic symmetry. Sample code: 515a. 

\textbf{MAPbI$_3$ crystals.}
Methylammonium lead triiodine (MAPbI$_3$) single crystals were low temperature solution-grown in a reactive inverse temperature crystallization (RITC) process, which utilizes a mixture of GBL  with alcohol~\cite{hocker2021}. The mixed precursor solvent polarity is changed compared to pure GBL, causing a lower solubility of MAPbI$_3$ and an optimization of nucleation rates and centers, which result in an early crystallization at low temperatures. Black MAPbI$_3$ single crystals were obtained at a temperature of 85$^\circ$C. At room temperature a tetragonal phase with lattice constants of $a=0.893$~nm and $c=1.25$~nm was determined by powder X-ray diffraction (XRD) in reflection geometry~\cite{hocker2021}. The size of the studied crystal is about $4 \times 3\times 2$~mm$^3$. The crystal has the shape of a planar truncated dodecahedron.  The front facet was X-ray characterized to point along the a-axis \cite{hocker2021}. Sample code: MAPI-SC04.

\textbf{FAPbBr$_3$ crystals.}
The FAPbBr$_3$ single crystals were grown with an analogous approach as the other samples following the ITC approach. Specific extended information are given in Ref.~\onlinecite{saidaminov2015}. The crystal is of reddish transparent appearance and has a rectangular cuboid shape with a size of $5 \times 5 \times 2$~mm$^3$. Sample code: OH0071a.

\textbf{CsPbBr$_3$ crystals.} 
The CsPbBr$_3$ crystals were grown with a slight modification of the ITC as stated above. Further information can be found in Ref.~\cite{Dirin2016}. First, CsBr and PbBr$_2$ were dissolved in dimethyl sulfoxide. Afterward a cyclohexanol in \textit{N,N}-dimethylformamide (DMF) solution was added. The resulting mixture was heated in an oil bath to $105^\circ$C whereby slow crystal growth appears. The obtained crystals were taken out of the solution and quickly loaded into a vessel with hot ($100^\circ$C) DMF. Once loaded, the vessel was slowly cooled down to about $50^\circ$C. After that, the crystals were isolated, wiped with filter paper and dried. The obtained rectangular-shaped CsPbBr$_3$ is crystallized in the orthorhombic modification. The crystals have one selected (long) direction along the $c$-axis [002] and two nearly identical directions along the $[\bar{1}10]$ and [110] axes~\cite{feng2020}. The size of the studied crystal is about $3\times2\times7$~mm$^3$. Sample code: DD4470/2.

\begin{figure*}[hbt]
\begin{center}
\includegraphics[width = 18cm]{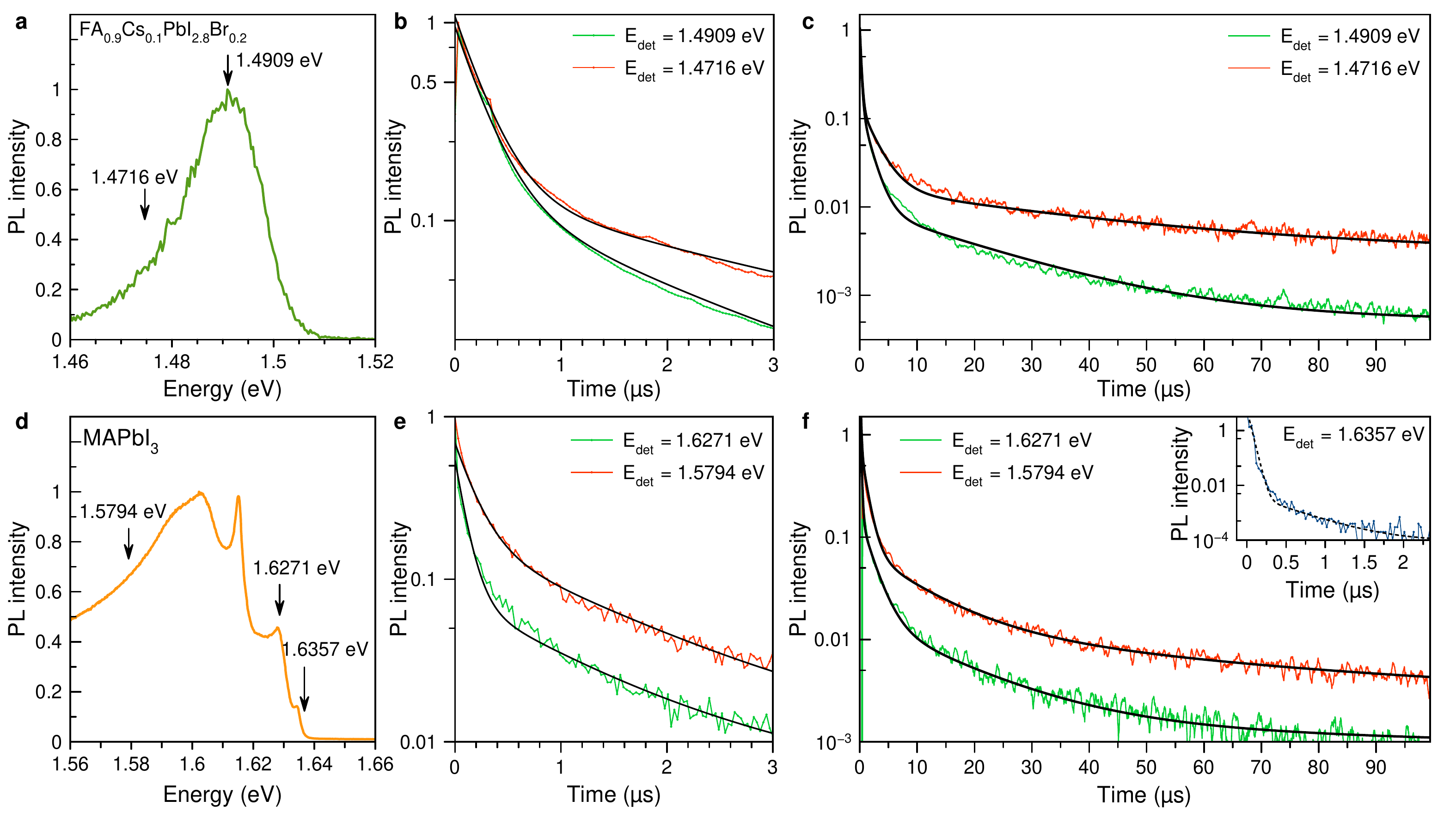}
\caption{\label{figS1} Time-integrated PL of (a) FA$_{0.9}$Cs$_{0.1}$PbI$_{2.8}$Br$_{0.2}$ and (d) MAPbI$_3$ measured at $T = 1.6$\,K. The arrows show the spectral positions at which the PL dynamics were detected. Time-resolved PL measured at various energies for (b),(c) FA$_{0.9}$Cs$_{0.1}$PbI$_{2.8}$Br$_{0.2}$ and (e),(f) MAPbI$_3$ in a 3~$\mu$s temporal range and in a 100~$\mu$s temporal range, respectively. The PL dynamics are fitted with a multi-exponential function with characteristic times collected in Table~\ref{tab:St1}. The fits are shown by the black lines.}
\end{center}
\end{figure*}

\textbf{MAPb(Br$_{0.05}$Cl$_{0.95}$)$_3$ crystals.}
2.2~mmol PbCl$_2$, 1.2~mmol MACl and 1~mmol MABr were dissolved in 2~ml of 1:1 DMF:DMSO mixture. The solution was filtered through an 0.45~$\mu$m polytetrafluoroethylene (PTFE) filter. The obtained solution was slowly heated in an oil bath up to 62$^\circ$C. The crystals nucleate and grow in the temperature window 58--62$^\circ$C. The obtained crystal has a cuboid shape with dimensions of $1.64 \times 1.65 \times 2.33$~mm$^3$. The crystal is transparent and colorless. Sample code: dd2924.

\subsection*{S2. Time-resolved photoluminescence}
\label{TRPL}

The exciton recombination times were measured by pump-probe differential reflectivity and by time-resolved photoluminescence with a streak camera. In FA$_{0.9}$Cs$_{0.1}$PbI$_{2.8}$Br$_{0.2}$ at $T=6$~K the decay of the exciton population dynamics is 0.45~ns~\cite{kirstein2021}. In MAPbI$_3$ at $T=4$~K it is 0.3~ns~\cite{kirstein2022mapi} and in  CsPbBr$_3$  at $T=10$~K it is 0.9~ns~\cite{belykh2019}. 

In this study, the long-lived recombination dynamics up to 100~$\mu$s were measured for the lead halide perovskite crystals by time-resolved photoluminescence (PL). The PL was excited by a pulsed laser with photon energy of 2.33~eV (532~nm wavelength), pulse duration of 5~ns, repetition rate of 10~kHz, and average excitation power of 8~$\mu$W. The detection energy ($E_\text{det}$) was selected by a double monochromator. The signal was detected using an avalanche photodiode and a time-of-flight card with a time resolution of 30~ns. 

Figure~\ref{figS1} shows the recombination dynamics in the FA$_{0.9}$Cs$_{0.1}$PbI$_{2.8}$Br$_{0.2}$ and MAPbI$_3$ crystals detected at various spectral energies. The time-integrated PL spectra measured for pulsed laser excitation are shown in Figure~\ref{figS1}a for FA$_{0.9}$Cs$_{0.1}$PbI$_{2.8}$Br$_{0.2}$ and in Figure~\ref{figS1}d for MAPbI$_3$. The arrows in the panels indicate the detection energies of the photoluminescence dynamics. The experiments were performed at the temperature of $T = 1.6$\,K. The recombination dynamics cover a broad temporal range up to 100\,$\mu$s. They cannot be fitted by monoexponential decays, evidencing that several recombination processes are involved. The PL dynamics contains three exponential decays for FA$_{0.9}$Cs$_{0.1}$PbI$_{2.8}$Br$_{0.2}$ in addition to the short exciton recombination time, which is too short to be resolved in this experiment, see Figures~\ref{figS1}b,c.
The dynamics of MAPbI$_3$ show four exponential decays, see Figures~\ref{figS1}e,f. The extracted times are given in Table~\ref{tab:St1}. 

The long recombination dynamics can be associated with the following processes: (i) recombination of electrons and holes localized at different crystal sites with significant dispersion in their separations~\cite{belykh2019,kirstein2022mapi,kirstein2021,kirstein2021nc}, (ii) carrier trapping and detrapping processes~\cite{Chirvony2018}, (iii) polaron formation~\cite{deQuilettes2019}, (iv) slow in-depth carrier diffusion~\cite{Bercegol2018}. The clarification of the role of the specific mechanisms goes beyond the scope of the present study. Spectral dispersion of the recombination times is observed for both samples. 

\begin{table*}[hbt]
\caption{Recombination times for FA$_{0.9}$Cs$_{0.1}$PbI$_{2.8}$Br$_{0.2}$ and MAPbI$_3$ at different spectral positions. $T = 1.6$\,K.}
\label{tab:St1}
\begin{center}
\begin{tabular*}
{0.55\textwidth}{@{\extracolsep{\fill}} |>{\centering\arraybackslash} m{0.1\textwidth}|>{\centering\arraybackslash} m{0.1\textwidth}|>{\centering\arraybackslash} m{0.1\textwidth}|>{\centering\arraybackslash} m{0.1\textwidth}|>{\centering\arraybackslash} m{0.1\textwidth}|}
\hline
$E_\text{det}$\,(eV) & $\tau_1$\,(ns)& $\tau_2$\,($\mu$s)& $\tau_3$\,($\mu$s) & $\tau_4$\,($\mu$s)\\
\hline
\multicolumn{5}{|c|}{FA$_{0.9}$Cs$_{0.1}$PbI$_{2.8}$Br$_{0.2}$}\\
\hline
1.4909 & 250 & 1.6 & 20 & - \\
\hline
1.4716 & 200 & 3.3 & 44 & -\\

\hline
\multicolumn{5}{|c|}{MAPbI$_3$}\\
\hline
1.6357 & 40 & 0.45 & - & - \\
\hline
1.6271 & 100 & 1.0 & 5.2 & 53 \\
\hline
1.5794 & 180 & 1.4 & 9.4 & 49 \\
\hline
\end{tabular*}
\end{center}
\end{table*}


\subsection*{S3. Anisotropy of electron and hole $g$-factors in FAPbBr$_3$}
\label{TRRE_aniso}

\begin{figure}[hbt]
\begin{center}
\includegraphics[width = 8.6cm]{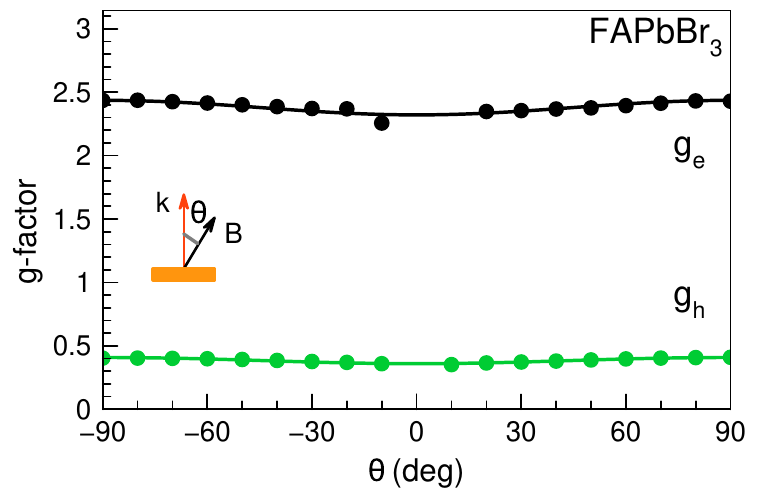}
\caption{\label{figS_TRKE} Angle dependence of electron and hole $g$-factors measured for FAPbBr$_3$ by time-resolved Kerr ellipticity using the experimental parameters given in the caption of Figure~2 of the main text.}
\end{center}
\end{figure}

We measured the anisotropy of the $g$-factors of electrons and holes in FAPbBr$_3$ by rotating the magnetic field in a vector magnet using time-resolved Kerr ellipticity. The magnetic field direction was changed from the Voigt geometry ($\theta = 90^\circ$) to the Faraday geometry ($\theta = 0^\circ$). The oscillations disappear in the Faraday geometry, but fitting by Eq.~(2) from the main text allows us to determine $g_\text{F,e}$ and $g_\text{F,h}$. The dependence of the $g$-factor values of electron and hole on $\theta$ is shown in Figure~\ref{figS_TRKE}. The anisotropy is small, namely, the $g$-factors in Faraday geometry $g_\text{F,e} = +2.32$ and $g_\text{F,h} = +0.36$ are close to those in Voigt geometry $g_\text{V,e} = +2.44$ and $g_\text{V,h} = +0.41$.

\subsection*{S4. Spin-flip Raman scattering in CsPbBr$_3$}
\label{SFRS_SI}

\begin{figure}[htb]
\begin{center}
\includegraphics[width = 8cm]{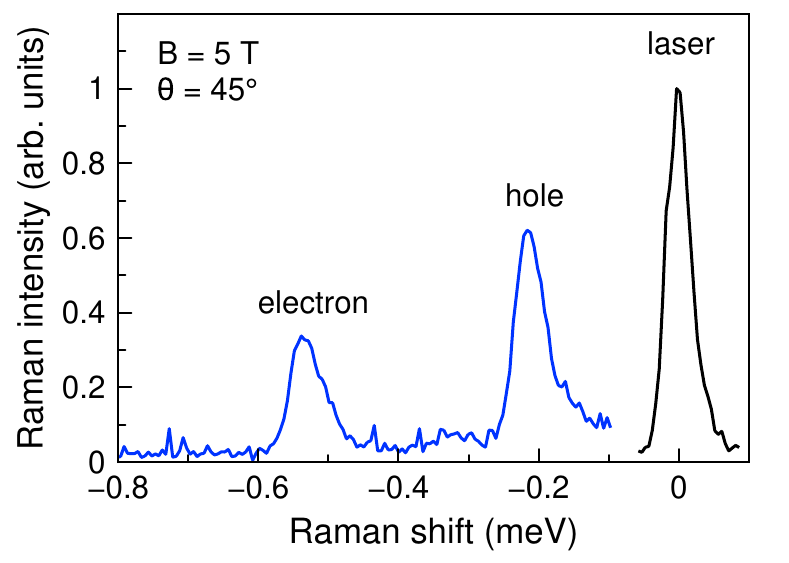}
\caption{\label{figS2} Anti-Stokes spin-flip Raman scattering spectrum of CsPbBr$_3$ measured at $T = 1.6$\,K. $\sigma^-$ polarization is used for excitation and $\sigma^+$ for detection. $E_\text{exc} = 2.331$\,eV, $B = 5$\,T, $\theta = 45^{\circ}$. }
\end{center}
\end{figure}

An example of spin-flip Raman scattering (SFRS) spectrum in CsPbBr$_3$ is given in Figure~\ref{figS2}, which was measured in tilted magnetic field geometry with $\theta = 45^{\circ}$. In order to reduce the background contribution of the resonantly excited photoluminescence, the SFRS spectrum was measured in the anti-Stokes spectral domain, i.e. the measured spectral range is shifted to higher energies from the laser. The laser excitation energy $E_\text{exc} = 2.331$\,eV was in the vicinity of the exciton resonance. Clearly resolved lines of the hole and electron spin-flips are seen. Their spectral shifts at the magnetic field of 5~T correspond to $g_\text{e}=+1.85$ and $g_\text{h}=+0.75$. 

\subsection*{S5. Formalism for anisotropic carrier and exciton $g$-factors}
\label{aniso_eq_SI}

Detailed model consideration of the electron, hole and exciton $g$-factors, including their anisotropy in the lead halide perovskites with tetragonal symmetry is present in Ref.~\cite{kirstein2021nc}. Here we give for convenience some equations from this paper. 

For the bottom conduction band one has
\begin{subequations}
\label{g:cb}
\begin{equation}
\label{g:cb:par}
\hspace{-16 mm}g_{e\parallel} = - \frac23+ \frac{2p_\perp^2}{m_0}  \frac{\cos^2\vartheta}{E_g},
\end{equation}
\begin{equation}
\label{g:cb:perp}
g_{e\perp} = - \frac23+ \frac{2\sqrt{2} p_\parallel p_\perp}{m_0} \frac{\cos{\vartheta} \sin{\vartheta} }{E_g}.
\end{equation}
\end{subequations}
Here the subscripts $\parallel$ and $\perp$ of the $g$-factors denote the direction of the magnetic field with respect to the $C_4$ axis (c-axis). $p_\parallel$ and $ p_\perp$ are the interband momentum matrix elements.
$\vartheta$ is the parameter that determines the relation between the crystalline splitting and the spin-orbit interaction.

For the valence band one obtains
\begin{subequations}
\label{g:vb}
\begin{equation}
\label{g:vb:par}
g_{h\parallel} = 2- \frac{2p_\perp^2}{m_0}  \left(\frac{\cos^2\vartheta}{E_g} +\frac{\sin^2\vartheta}{E_g+\Delta_{le}} - \frac{1}{E_g + \Delta_{he}}\right),
\end{equation}
\begin{equation}
\label{g:vb:perp}
g_{h\perp} = 2- \frac{2\sqrt{2} p_\parallel p_\perp}{m_0} \cos{\vartheta} \sin{\vartheta} \left(\frac{1}{E_g} - \frac{1}{E_g+\Delta_{le}}  \right).
\end{equation}
\end{subequations}
Here, $\Delta_{he}$ and $\Delta_{le}$ are split-orbit splittings to between the bottom conduction band and the higher-energy bands of heavy electrons (he) and light electrons(le). In the cubic case $\Delta_{le} = \Delta_{he}\equiv \Delta$ and $p_\parallel = p_\perp \equiv p$.

Note that the valence band $g$-factors in the electron and hole representation have the same sign because the transformation from the electron to the hole representation includes both a change in the sign of energy and the time reversal. We define the Land\'e factor in such a way that, e.g., for $\bm B \parallel z$ the splitting $E_{+1/2} - E_{-1/2}$ between the states with spin projection $+1/2$ and $-1/2$ onto the $z$ axis is given by $g_{e\parallel} \mu_B B_z$ and $g_{h\parallel} \mu_B B_z$.

Using Eqs.~\eqref{g:cb} and \eqref{g:vb} we evaluate also the bright exciton $g$-factor, which describes the splitting of the exciton radiative doublet into circularly polarized components, by $g_X=g_e+g_h$. In the tetragonal crystals the exciton $g$-factor is
\begin{subequations}
\label{g:x}
\begin{equation}
\label{g:x1}
g_{X\parallel} = g_{e\parallel}+ g_{h\parallel} = {\frac43} - \frac{2p_\perp^2}{m_0}  \left(\frac{\sin^2\vartheta}{E_g+\Delta_{le}} - \frac{1}{E_g + \Delta_{he}}\right) + \Delta g_{e},
\end{equation}
\begin{equation}
\label{g:x2}
g_{X\perp} = g_{e\perp} + g_{h\perp} = {\frac43} + \frac{2\sqrt{2} p_\parallel p_\perp}{m_0}  \frac{\cos{\vartheta} \sin{\vartheta}}{E_g+\Delta_{le}}  + \Delta g_{e}.
\end{equation}
\end{subequations}
One can see, that the contributions to the individual $g$-factors due to the $\bm k\cdot\bm p$-mixing of the valence band with the bottom conduction band $\propto 1/E_g$  cancel each other in the exciton $g$-factor.

\subsection*{S6. Modeling of the dark exciton $g$-factor}
\label{Dark_X_SI}

\begin{figure}[htb]
\begin{center}
\includegraphics[width = 8cm]{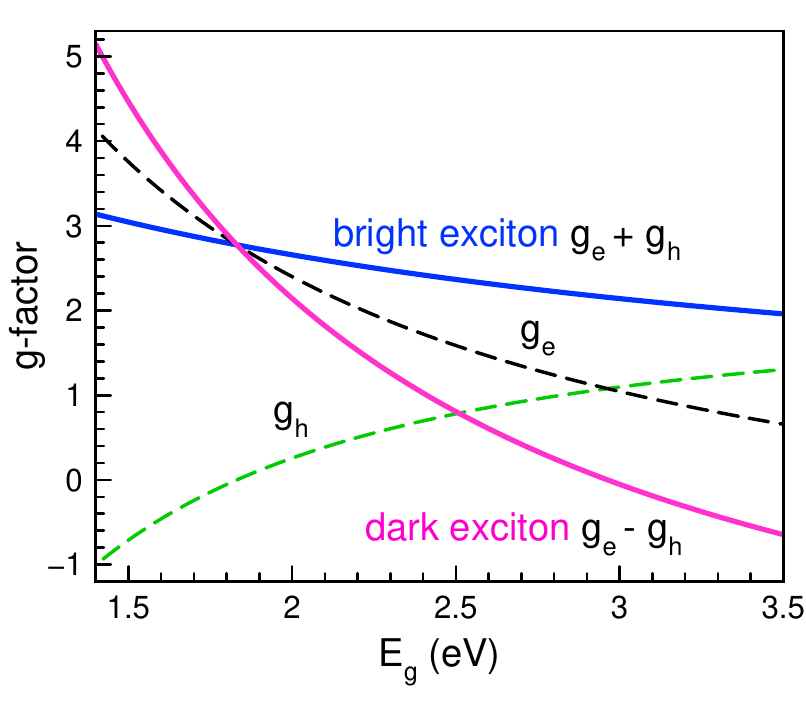}
\caption{\label{figS4} Modeling of the $g$-factor band gap energy dependence for the bright and dark excitons in lead halide perovskites. Dashed lines are model calculations with Eqs.~(3) and (4) from the main text: for the hole (green) and electron (black) $g$-factors. Blue line is for the bright exciton, calculated as $g_\text{e}+g_\text{h}$. Purple line is for the dark exciton, calculated as $g_\text{e}-g_\text{h}$. }
\end{center}
\end{figure}

Here we give model results for the dark exciton $g$-factor calculated as $g_\text{e}-g_\text{h}$. We show in Figure~\ref{figS4} model calculations for the electron, hole and bright exciton, which are presented in Figure~4 in the main text, and add here results for the dark exciton (purple line). One can see that the dark exciton dependence is stronger than the one for the bright exciton and even stronger than the dependence for the electron.

\newpage
\pagebreak